\documentclass{aastex}
\usepackage{spr-astr-addons}
\usepackage{url}

\RequirePackage{color}

\begin{document}

\title{Some New Possible Anticipated Signals for Existence of Magnetic Monopoles} \slugcomment{Not to appear in
Nonlearned J., 45.}
\shorttitle{A Possible Influence on Standard Model}
\shortauthors{Qiu-He Peng et al.}

\author{Qiu-He Peng \altaffilmark{1}, Jing-Jing, Liu \altaffilmark{2}, and Zhong-Qi, Ma\altaffilmark{3}}
\affil{Corresponding to Jing-Jing, Liu}
\email{liujingjing68@126.com}

\altaffiltext{1}{Department of Astronomy, Nanjing University,
Nanjing, Jiangshu 210000, China.}
\altaffiltext{2}{College of Marine
Science and Technology, Hainan Tropical Ocean University, Sanya,
Hainan 572022, China.}

\altaffiltext{3}{Institute of High Energy Physics, Chinese Academy
of Sciences, Beijing 100049, China.}

\begin{abstract}
We summarize some predictions from the model of supermassive object
with magnetic monopoles which match up with recent astronomical
observations quantitatively. They may be the signals for existence
of magnetic monopoles in the supermassive objects, such as one at
the Galactic Center.
\end{abstract}

\keywords{Galaxy: center --- Galaxies: magnetic fields --- Physical
Date and Processes: black hole physics.}


\section{Introduction}
\label{sect:intro} Most of physicists believe that the existence of
magnetic monopoles had been ruled out by experiments. However,
experiments only indicated that the flux of magnetic monopoles on
the earth is too low to be observed. In the primordial universe, the
electromagnetic interaction between magnetic monopoles and plasma is
so strong such that magnetic monopoles might store up in the center
of quasars and active galactic nuclei (AGN) during the collapsing
process of the original giant nebulae, including at the collapsed
core of the Galactic Center (GC). Due to the Rubakov-Callan (RC)
effect\citep{rc,rc2}, magnetic monopoles may catalyze nucleon decay
and is invoked as the energy source of quasars and AGN. The model of
supermassive object with magnetic monopoles (SMOMM) \citep{peng},
presented by one (QP) of the authors, gave five predictions, some of
which match up with the recent astronomical observations
quantitatively. It may be the signals for existence of magnetic
monopoles in the supermassive objects.

From the symmetry and beauty of the Dirac equation, Dirac
\citep{dirac} suggested the existence of magnetic monopoles which is
the most natural explanation of the quantization of electric charge.
It is a fundamental problem in physics whether magnetic monopoles
exist or not.

The plan of this letter is as follows. In section 2, we will sketch
the model of SMOMM and its predictions. In Sections $3\sim 5$, we
will show some predictions of our model in agreement quantitatively
or basically with some new astrophysical observations: a strong
magnetic field around the GC, the observation of the rate of the
emitted positrons, and the frequency of the spectrum peak of the
thermal radiation from the GC. In the last Section, we will discuss
and try to explain two recent observations with the model of the
stars with the magnetic monopoles: the highly super-luminous
supernova ASASSN-15lh and the gravitational waves GW150914.

\section{The model of SMOMM}
\label{sect:The}

The model of SMOMM \citep{peng} suggested that an amount of magnetic
monopoles stored up in the center of quasars and AGNs, including at
the collapsed core of the GC during the collapsing process of the
original giant nebulae in the primordial universe because the
electromagnetic interaction between magnetic monopoles and plasma
was so strong. Due to the RC effect \citep{rc,rc2}, nucleon decay
may be catalyzed by magnetic monopoles,
\begin{equation}
p+M \rightarrow~M+e^{+}+\pi^{0}+{\rm debris} \ (85\%),
\end{equation}
\begin{equation}
p+M\rightarrow~M+e^{+}+\mu^{+}+\mu^{-}+{\rm debris} \ (15\%).
 \label{2}
\end{equation}

\noindent In this model, the RC effect is invoked as the energy
source. In the case of GC, the SMOMM is located at the center with
the radius of about $8.1\times 10^{15}cm \sim 1.2\times 10^{4}
R_{S}$ ($R_{S}$ is its Schwarzschild radius). The total mass of the
SMOMM is derived to be $2.5 \times 10^{6} M_{\bigodot}$ (now taken
as $4.6 \times 10^{6} M_{\bigodot}$) from the observed luminosity of
Sgr A$^{*}(1\times 10^{37}$ ergs ${\rm s}^{-1}$).

The gravitational effect around SMOMM in the GC is similar to the
model of supermassive black holes. In the black hole model,
accretion of matter results in the huge luminosity, but the energy
source in the model of SMOMM is supplied by the RC effect.

\vspace{2mm} The main predictions of the model of SMOMM are as
follows \citep{peng}.

\vspace{2mm} 1. The production rate of positrons emitted from the
SMOMM in the model is $\sim 6.5\times 10^{42}e^{+}/sec$.

\vspace{2mm}

2. High-energy gamma-ray radiation has energy higher than 0.5 MeV.
The integrated energy of these radiations would be much greater than
both the bolometric luminosity and the energy of positron
annihilation line.

\vspace{2mm}

3. The radial magnetic field at the surface of the SMOMM is
estimated to be $H(R)\sim (20-100) G$.

\vspace{2mm}

4. The strong radial magnetic fields of the high-speed rotating
SMOMM transforms a strong electric field for a distant observer in
the rest frame. A variety of produced particles ($e^{+}$,
$\mu^{\pm}$, $\pi^{\pm}$) would be accelerated by the strong
electric field to very high energy, say $E_{\gamma}\sim 10^{21}$eV
or greater. We predict that these could just be the observed
ultra-high-energy cosmic rays which have an initial energy of
several hundred MeV produced from the SMOMM.

\vspace{2mm}

5. The surface temperature of the SMOMM is derived to be about 121 K
and the corresponding spectrum peak of the thermal radiation is at
$10^{13} Hz$ in the sub-mm wavelength regime.

\vspace{3mm}

\section{A strong magnetic field around the galactic center}

The recent observation \citep{mag} in 2013 indicated that {\bf there
is a dynamically important magnetic field near the black hole}. In
particular, at $r=0.12 pc$ the lower limit of the outward radial
magnetic field near the GC is
\begin{equation}
B\geq 8\left[\displaystyle \frac {RM}{66.960 m^{-2}}\right]
\left[\displaystyle \frac {n_{0}}{26 cm^{-3}}\right]^{-1} mG,
 \label{3}
\end{equation}

\noindent where $n_{0}$ is the number density of electrons there,
and RM denotes the measurement of the Faraday rotation near the GC.
The lower limit of the observed data is in agreement with the the
prediction 3 in the model of SMOMM because the magnetic field
strength decreases as the inverse square of the distance from the
source and has $B\approx (10-50)mG$ at $r=0.12 pc$. Up to now no
other physical mechanism can produce this strong radial magnetic
field.

As analyzed in \citep{zam}, ``jet magnetic field and accretion disk
luminosity are tightly correlated over seven orders of magnitude for
sample of 76 radio-loud active galaxies". They pointed out that the
black hole models ``may require significant changes", and ``models
of the Galactic Center accretion disk may also need to be revised,
as a dynamically important magnetic field has been reported
\citep{mag} within a distance of $\sim 3\times 10^{7} r_{g}$ from
the central black hole."

\section{Rate of emitted positrons}

New observation \citep{kno} reported that the measured 511 keV line
flux located at the GC at a distance of $8.5kpc$ converts into an
annihilation rate of $(3.4-6.3)\times 10^{42} s^{-1}$. ``The
observed flux is compatible with previous measurements
\citep{sha,che,pur,mil,mil2} that have been obtained using
telescopes with small or moderate fields-of-view, yet it is on the
low side when compared to OSSE measurements." \citep{kno} Those
observations are in agreement with the the prediction 1 in the model
of SMOMM quantitatively.

\section{Frequency of spectrum peak of the thermal radition from the galactic center}

A review paper \citep{fal} pointed out that the radio flux density
$S_{\nu}$ from the GC shows a flat-to-inverted spectrum. i.e., it
raised slowly with frequency of the power peaking around $10^{12}$
Hz in the sub-mm band. The observed power peak is in agreement
basically with the prediction 5 in the model of SMOMM.

\section{Conclusions and discussions}
\label{sect:conclusion}

The agreement of the predictions of our model of the SMOMM with
three new astrophysical observations quantitatively or basically
issues the signals for existence of magnetic monopoles. We are
looking forward to seeing more astrophysical observation which will
meet the predictions of our model.

At the beginning of this year two important astrophysical
observations were reported: The highly super-luminous supernova
ASASSN-15lh and the gravitational waves GW150914. We believe that
our model and its development can be suitable to meet the new
observations.

The recent observation \citep{don} reported the discovery of
ASASSN-15lh (SN 2015L), which was interpreted as the most luminous
supernova to be found. ``At redshift $z-0.2326$, ASASSN-15lh reached
an absolute magnitude of $M_{u,AB}=-23.5\pm 0.1$ and bolometric
luminosity $L_{bol}= (2.2\pm 0.2)\times 10^{45}$ erg s$^{-1}$, which
is more than twice as luminous as any previously known supernova".
``In the 4 months since first detection, ASASSN-15lh radiated
$(1.1\pm 0.2)\times 10^{52}$ ergs, challenging the magnetar model
for its engine".

Up to now, the supernova explosion mechanism has not been solved. In
a model of a supermassive star with magnetic monopoles, the
phenomena of ASASSN-15lh and the super-luminous supernova (SLSN) may
be explained naturally. Although a few magnetic monopoles may be
stored in the core of stars during the process of star formation
from a neutral hydrogen nebulae due to very weak interaction between
magnetic monopoles and neutral hydrogens, the stars may capture some
flight magnetic monopoles in the space \citep{pen}. The flux of the
flight magnetic monopoles is
\begin{eqnarray}
\Phi_{m}=n_{m}v_{m}=10^{-4}cn_{B}^{(0)}\zeta^{(0)}_{m}(\displaystyle \frac{v_{m}}{10^{-4}c})\nonumber\\
 \approx 7.5\times 10^{-19}(\displaystyle \frac{\zeta^{(0)}_{m}}{\zeta_{S}})(\displaystyle \frac{n_{B}^{(0)}}{1cm^{3}})(\displaystyle \frac{v_{m}}{10^{-4}c})cm^{-2}s^{-1}\nonumber\\
 \approx 2.5\times 10^{-4}(\displaystyle \frac{\zeta^{(0)}_{m}}{\zeta_{S}})(\displaystyle \frac{n_{B}^{(0)}}{1cm^{3}})(\displaystyle \frac{v_{m}}{10^{-4}c})\times (100m)^{-2}Yr^{-1},
 \label{4}
\end{eqnarray}

\noindent where $n_m$ is the number density of the magnetic
monopoles, $v_{m}$ is the average velocity of the flight magnetic
monopoles in space, $n_{B}^{(0)}$ is the number density of the
baryons in the interstellar space, and
$\zeta_{S}=Gm_{B}m_{m}/g_{m}^{2} \simeq 1.9\times 10^{-25}$ denotes
the Newton saturation value of $\zeta_m=N_{m}/N_{B}$, where $N_m$ is
the total number of the flight magnetic monopoles and $N_B$ is the
total number of nucleons.
%
$\zeta_m^{(0)}$ is the value of $\zeta_m$ in the interstellar space,
and its upper limit is $\zeta_m^{0}\leq 10^{-20\pm 1}$
\citep{par,laz}.  After formation of stars the total number of the
magnetic monopoles captured by the stars from space is estimated to
be

\begin{eqnarray}
N_{m}&=4\pi R^{2}\Phi_{m}T\approx 3\times 10^{24}\left(\displaystyle
\frac{\zeta^{(0)}_{m}}{\zeta_{S}}\right)
\left(\displaystyle \frac{n_{B}^{(0)}}{1cm^{3}}\right)\nonumber\\
&~~~\times \left(\displaystyle \frac{v_{m}}{10^{-4}c}\right)
\left(\displaystyle \frac{R}{10^{3}R_{\bigodot}}\right)^{2}
\left(\displaystyle \frac{T}{10^{7}Yr}\right).
 \label{5}
\end{eqnarray}

\noindent where $R$ is the radius of the star, $T$ is the age of the
progenitor of the supernova.
The captured superheavy monopoles are gathered at the core of the
stars.

Due to the RC effect, the luminosity produced by the nucleon decay
catalyzed by magnetic monopoles in the core of the stars is
\begin{equation}
L_{m}=N_{m}\langle \sigma v\rangle n_{B}^{(c)}m_{B}c^{2},
 \label{6}
\end{equation}

\noindent where $\sigma\sim 10^{-25}-10^{-26}$ cm$^{2}$ is the cross
section of the RC effect \citep{rc,rc2}, $v$ is the thermal velocity
of the nucleons, and $n_{B}^{(c)}$ is the central number density of
the baryons for the star.
The temperature would reach $10^{11}$K in the collapsed core of the
supernova, i.e., $v/c\geq 0.1$. Thus,

\begin{eqnarray}
L_{m}&\sim (10^{42}-10^{43}) \left(\displaystyle
\frac{n_{B}^{(c)}}{n_{nuc}}\right)
\left(\displaystyle \frac{\sigma}{10^{-26}cm^{2}}\right)\nonumber\\
&~~~\times \left(\displaystyle \frac{R}{10^{3}R_{\bigodot}}\right)^2
\left(\displaystyle \frac{T}{10^{7}Yr}\right).
\label{7}
\end{eqnarray}

\noindent This luminosity is enough to explain the supernova
explosion, when the density of the supernova core is greater than
the nuclear density ($n_{nuc}$).

As long as we assume that the initial mass of the progenitor of the
SLSN is greater than $150 m_{\bigodot}$, then its radius before the
supernova explosion was greater than $10^{4} R_{\bigodot}$, and the
life of the super-massive star is only $10^{5}-10^{6} Yr$. Thus, the
luminosity produced from the nucleon decay catalyzed by the magnetic
monopoles in the collapsed core of the supernova, where the density
of the core is much greater than the nuclear density, will be
greater than $10^{42}-10^{43}$ ergs. This luminosity is enough to
make the SLSN exploding, so that the SLSN is explained naturally.

On September 14, 2015 two detectors of the Laser Interferometer
Gravitational-Wave Observatory (LIGO) simultaneously observed a
transient gravitational-wave signal (GW150914) \citep{ligo}. This is
the first direct detection of gravitational waves which was commonly
explained as the first observation of a binary black hole system
mergering to form a single black hole.  On December 26, 2015, a
second gravitational-wave event, GW151226, was observed by the twin
detectors of LIGO\citep{ligo2}.

On February 16, 2016, \citet{con} reported that the Fermi Gamma-ray
Burst Monitor (GBM) had revealed the presence of a weak transient
source above 50 keV, 0.4 s after the GW event: ``This weak transient
lasting 1 s does not appear connected with other previously known
astrophysical, solar, terrestrial, or magnetospheric activity. Its
localization is ill-constrained but consistent with the direction of
GW150914''\citep{con}. They also note that the electromagnetic
signal from a stellar mass black hole binary merger is not expected
if the GBM transient is associated with GW150914.
Based on their measurement of the fluence seen by GMB, a luminosity
of $1.8^{+1.5}_{-1.0} \times 10^{49}$ erg s$^{-1}$ is derived in
hard X-ray emission between 1 keV and 10 MeV.

In another paper \citep{lyu} Lyutikov analyzed the physical
requirements in some detail that the possible observation of the
electromagnetic signal contemporaneous with GW150914 imposes on the
circum-merger environment, and found that the required physical
parameters at the source exceed by many orders of magnitude what is
expected in realistic astrophysical scenarios. He concluded that
Fermi GMB signal contemporaneous with GW150914 is unrelated to the
black hole merger.

The merger of two black holes can produce the GW only, not the
gamma-ray burst simultaneously.  However, in our model of stars with
magnetic monopoles, the merger of two supermassive neutron stars
with the magnetic monopoles, (They cannot collapse to black holes
through the RC effect), whose mass density reaches to
$(10^{5}-10^{6})\rho_{nuc}$ at the center core, not only produces
the GW, but also produces the gamma-ray burst whose luminosity may
reach $10^{49} ergs/s$ through the RC effect.
After we have finished this paper, \citet{Racusin} reported on June
15, 2016 that they had made observations of the event GW151226 and
candidate LVT151012 with both the Fermi Gamma-ray Burst Monitor and
the Large Area Telescope (LAT).  No electromagnetic counterparts
were detected by either GBM or LAT.
We are looking forward to the future detections of the GW which may
accompany with the gamma-ray burst. If yes, it will verify the
existence of magnetic monopoles.

\acknowledgments We would like to thank Prof. Daniel Wang, Prof.
Y.F. Huang, Prof. P.F. Chen and Prof. J. L. Han for their help to
inform us some new information of observations. This work was
supported in part by the National Natural Science Foundation of
China under grants 10773005, 11565020, and the Counterpart
Foundation of Sanya under grant 2016PT43, the Special Foundation of
Science and Technology Cooperation for Advanced Academy and Regional
of Sanya under grant 2016YD28, and the Natural Science Foundation of
Hainan province under grant 114012.


\begin{thebibliography}{}
\bibitem[Abbott et al.(2016)]{ligo} Abbott, B.~P., Abbott, R., Abbott, T.~D., et al., 2016, Physical Review Letters, 116, 061102
\bibitem[Abbott et al.(2016)]{ligo2} Abbott, B.~P., et al., 2016, Phys. Rev. Lett., 116, 241103
\bibitem[Callan(1983)]{rc2} C. Callan, 1983, Nucl. Phys., 212, 391
\bibitem[Cheng et al.(1997)]{che} Cheng, L.~X., Leventhal, M., Smith, D.~M., et al., 1997, ApJL, 481, 43
\bibitem[Connaughton et al.(2016)]{con} Connaughton, V., Burns, E., Goldstein, A., et al.\ 2016, arXiv:1602.03920
\bibitem[Dirac(1958)]{dirac}  P. A. M. Dirac., 1958, The Principle of Quantum Mechanics, (Clarendon Press, Oxford, 1958).
\bibitem[Dong et al.(2016)]{don} Dong, S., Shappee, B.~J., Prieto, J.~L., et al., 2016, Science, 351, 257
\bibitem[Eatough et al.(2013)]{mag}  R.P. Eatough,  H. Falcke,  R. Karuppusamy et al., 2013, nature, 591, 391
\bibitem[Falcke \& Markoff(2013)]{fal} Falcke, H., \& Markoff, S.~B., 2013, Classical and Quantum Gravity, 30, 244003
\bibitem[Kn{\"o}dlseder et al.(2003)]{kno} Kn{\"o}dlseder, J., Lonjou, V., Jean, P., et al., 2003, Aap, 411, 457
\bibitem[Lazarides et al.(1981)]{laz} Lazarides, G., Shafi, Q., \& Walsh, T.~F., 1981, Physics Letters B, 100, 21
\bibitem[Lyutikov(2016)]{lyu} Lyutikov, M., 2016, arXiv:1602.07352
\bibitem[Milne et al.(2000)]{mil} Milne, P.~A., Kurfess, J.~D., Kinzer, R.~L., Leising, M.~D., \& Dixon, D.~D., 2000, American Institute
of Physics Conference Series, 510, 21
\bibitem[Milne et al.(2001)]{mil2} Milne, P.~A., Kurfess, J.~D., Kinzer, R.~L., \& Leising, M.~D., 2001, Gamma-Ray Astrophysics, 587, 11
\bibitem[Parker(1970)]{par} Parker, E.~N., 1970, ApJ, 160, 383
\bibitem[Peng et al.(1985)]{pen}  Peng, Q., Lie, Z. and Wang, D., 1985, Scientia Sinica (Serries A), 28, 970
\bibitem[Peng \& Chou(2001)]{peng}  Peng, Q. \&  Chou, C., 2001, ApJ, 551, 23
\bibitem[Purcell et al.(1997)]{pur} Purcell, W.~R., Cheng, L.-X., Dixon, D.~D., et al., 1997, ApJ, 491, 725
\bibitem[Racusin et al.(2016)]{Racusin} Racusin, J.~L., Burns, E., Goldstein, A., et al., 2016, arXiv:1606.04901
\bibitem[Rubakov(1981)]{rc}  Rubakov, V. 1981, JETP Lett. 33,644
\bibitem[Share et al.(1990)]{sha} Share, G.~H., Leising, M.~D., Messina, D.~C., \& Purcell, W.~R., 1999, ApJL, 358, 45
\bibitem[Zamaninasab et al.(2014)]{zam} Zamaninasab, M., Clausen-Brown, E., Savolainen, T., \& Tchekhovskoy, A., 2014, Nature, 510, 126

\end{thebibliography}
\end{document}